\documentclass{PoS}

\usepackage{amsmath}
\usepackage[caption=false]{subfig}

\title{Theory of CP violation in B decays}

\ShortTitle{Theory of CP violation in B decays}

\author{\speaker{K. Keri Vos} \\
           Theoretische Physik 1, Naturwissenschaftlich-Technische Fakult\"at, \\
           Universit\"at Siegen, D-57068 Siegen, Germany\\
       E-mail: \email{keri.vos@uni-siegen.de}}

\abstract{The study of CP violation in $B$-meson decays has already reached a high level of precision, which will be pushed even higher in the future era of Belle-II and the LHCb upgrade. %Important probes of CP violation are the $B_d^0$ and $B_s^0$ mixing phases and the CKM angle $\gamma$. 
Here, the theoretical challenge is to control the uncertainties from strong interactions to distinguish between the Standard Model and possible New Physics effects. 
In this talk, I will present a selected overview of recent theoretical developments in this field. This includes, in particular, the  semileptonic asymmetry $a_{\text{sl}}^s$ and extractions of the CKM angle $\gamma$ and the $B_d^0$ and $B_s^0$ mixing phases. I focus on recently proposed strategies in which the theory uncertainties can be controlled through data using flavour symmetries of the strong interaction. A newly found puzzle in the $B \to \pi K$ system is highlighted and a theoretically clean way to determine the underlying electroweak penguin parameters is discussed.  Finally, the recent progress to describe three-body $B$ decays in QCD factorization is discussed.}

\FullConference{The International Conference on B-Physics at Frontier Machines - BEAUTY2018\\
	6-11 May, 2018\\
	La Biodola, Elba Island, Italy}

\begin{document}
\section{Introduction}
Studies of CP violation are an important part of the flavour physics program at LHCb and the B factories. They are particularly interesting since they test the Standard Model of particle physics (SM) and might reveal new physics (NP). Within the SM, CP violation is described by the Cabibbo-Kobayashi-Maskawa (CKM) matrix. The main theoretical challenge is to disentangle the effects of new physics and strong interaction effects within the SM. Thanks to a combined theoretical and experimental endeavour, a impressive level of precision was already reached. At Belle-II \cite{BelleII} and the LHCb upgrade \cite{LHCb-up}, this precision will even be pushed to a much higher level and fully exploiting requiring a continued interplay between theory and experiment. 
 
In studies of CP violation, non-leptonic $B$ decays are the key players. The theoretical analyses of these decays are in general challenging due to hadronic matrix elements of four-quark operators entering the corresponding low-energy effective Hamiltonians. However, the flavour symmetries of strong interactions imply relations
between the different non-leptonic decays, thereby allowing the elimination of hadronic amplitudes or their determination from experimental data. 

In this talk, I focus on the recent theoretical progress in the study of CP violation using flavour symmetries. In particular, I will present newly proposed strategies in which theoretical uncertainties can be controlled using experimental data. This leads to a theoretical precision that matches the (expected) experimental precision. First, in Sec.~\ref{sec:mix}, we discuss the mixing angles $\phi_d$ and $\phi_s$ which probe CP violation in neutral $B_d^0$ and $B_s^0$ meson mixing, respectively. These phases can be determined from the semileptonic asymmetry $a_{\rm sl}$, but also from {\it exclusive} non-leptonic decays using the mixing-induced CP asymmetries. We discuss the interplay between these different determinations and the room for new physics. In Sec.~\ref{sec:nonlep}, we further discuss the non-leptonic decays, and in particular the determinations of the CKM parameter $\gamma$ and the mixing phases $\phi_s$ and $\phi_d$. Besides the determination of $\phi_s$ from tree decays, we discuss in Sec.~\ref{sec:BstoKK} a recent strategy to extract $\phi_s$ from the penguin dominated $B_s^0\to K^-K^+$ decay and its $U$-spin partner $B_d^0 \to \pi^- \pi^+$. Then we focus on the $B \to \pi K$ system, where a new puzzle is found using an isospin amplitude relation. This might hint at New Physics (NP) entering through the electroweak penguin topologies. A strategy to determine these parameters is discussed, which offers exciting prospect for the $B_d^0\to \pi^0 K_S$ CP asymmetries to reveal possible NP. Finally, we end with a brief discussion of the recent developments in three-body decays and some concluding remarks.   

\section{CP violation in $B_q-\bar{B}_q$ mixing}\label{sec:mix}
\begin{figure}[t]
\centering
\subfloat[]{\includegraphics[width=0.3\textwidth]{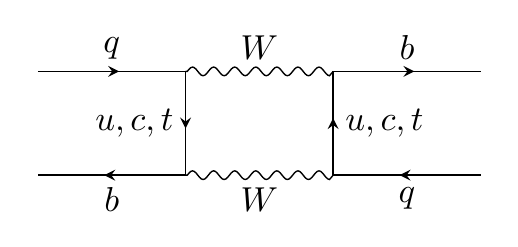}}
\subfloat[]{
\includegraphics[width=0.3\textwidth]{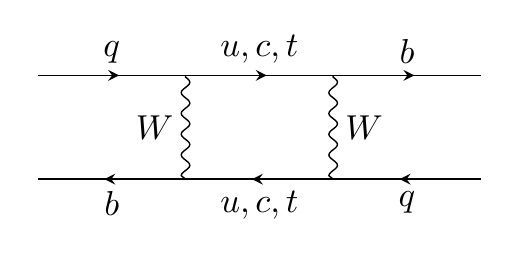}}
\caption{Leading contribution to $B_q$-$\bar{B}_q$ mixing in the SM. } 
\label{fig:boxes}
\end{figure}

In the SM, $B_q^0-\bar{B}_q^0$ mixing is governed by the box diagrams in Fig.~\ref{fig:boxes}. Because of this phenomenon, the initial $B_q^0$ meson evolves into a time-dependent linear combination of the $B_q^0$ and $\bar{B}_q^0$ states and can be described by a Schr\" odinger equation. Solving the Schr\" odinger equation then gives the physical mass eigenstates $H$ and $L$ 
\begin{equation}
\left|B_L\right\rangle = p \left|B_q^0\right\rangle + q \left|\bar{B}_q^0\right\rangle \;\; \rm{and}\;\; \left|B_H\right\rangle = p\left|B_q^0\right\rangle - q\left|\bar{B}_q^0\right\rangle \ ,
\end{equation}
with the corresponding masses  $M_H^q, M_L^q$ and the decay widths $\Gamma_H^q$ and $\Gamma_L^q$. This introduces three observables. First, the mass difference
\begin{equation} \Delta M_q \equiv M^q_H - M^q_L \sim 2 |M^q_{12}|> 0 \ ,
\end{equation}
which is dominated by short-distance contributions such that NP can have a significant impact \cite{Luz18}. On the other hand, the width difference
\begin{equation}
\Delta \Gamma_q \equiv \Gamma^q_L - \Gamma^q_H \sim 2 \Gamma^q_{12} \cos\phi_q 
\end{equation}
is governed by tree level contributions and is therefore expected to be rather insensitive to NP contributions \cite{Dun00, Bad07, Lenz-12}. For the $B_s$ system, the Particle Data Group (PDG) \cite{PDG} gives the averages 
\begin{equation}\label{eq:GammaMs}
\frac{\Delta \Gamma_s}{\Gamma_s}=0.124\pm0.011, 
\quad
\frac{\Delta M_s}{\Gamma_s}=26.81\pm0.10,
\end{equation}
where $1/\Gamma_s=(1.510\pm0.005)\times10^{-12}\,\mbox{s}$ is the $B^0_s$ lifetime. % $\Delta\Gamma_s$ is sizeable. 
Finally, CP violation in mixing gives rise to the mixing phase
\begin{equation}
\phi_q \equiv \rm{arg} \left(- M_{12}^q/\Gamma_{12}^q\right) \ .
\end{equation}

This phase can be probed using flavour-specific semileptonic decays via \cite{Dun00}
\begin{equation}\label{eq:asl}
a^q_{\rm{sl}} %= a^q_{\rm{fs}} 
= \frac{\Gamma(\bar{B}_q(t) \to f) - \Gamma(B_q(t) \to \bar{f})}{\Gamma(\bar{B}_q(t) \to f) + \Gamma(B_q(t) \to \bar{f})} = %\left|\frac{\Gamma^q_{12}}{M^q_{12}}\right| \sin\phi^q_{12} = 
							\left( \frac{\Delta\Gamma_q}{\Delta M_q}\right)\tan \phi_q \ ,
\end{equation}
where the direct transitions $\bar{B}_q(t) \to f$ and $B_q(t) \to \bar{f}$ are forbidden and can only occur via mixing. Examples of such flavour-specific decays are $B_s^0\to D_s^- \pi^+$ and $B_s^0 \to X \ell \bar\nu_\ell$. In the SM, the flavor-specific CP asymmetries of $B_d^0$ and $B_s^0$ can be determined using the heavy-quark expansion and inputs from Lattice QCD \cite{Art15}:
\begin{equation}
a_{\rm{sl}}^d|_{\rm{SM}}= (-4.7\pm 0.6)\times 10^{-4} \ , \quad a_{\rm{sl}}^s|_{\rm{SM}} = (2.22\pm 0.27)\times 10^{-5} \ .
\end{equation}
We note that this prediction assumes quark-hadron duality \cite{Jub16}. Due to the small SM values, especially for $a_{\rm{sl}}^s$, any sizeable experimental deviation from this prediction would be a clear sign of new physics. However, such CP-violating NP would also affect the mixing-induced CP asymmetry of non-leptonic $B$ decays. 

\subsection{A closer look at $a^s_{\rm sl}$}
Let us now focus on the $B_s-\bar{B}_s$ system and the constraints on $a^s_{\rm sl}$ from measurements of CP violation in {\it exclusive} $B$ decays \cite{KKV17}. For $\bar{B}_s^0 \to f$ decays with a final state $f = J/\psi \phi, D_s^- D_s^+,  J/\psi \pi^+\pi^-$ caused by $b\to c\bar{c}s$ processes, measurements of the CP asymmetries allow the extraction of 
\begin{equation}\label{eq:phisfdef}
\phi_s^f = \phi_s^{\text{SM}} + \phi_s^{\text{NP}} + \Delta \psi^{\text{SM}}_f +\Delta \psi^{\text{SM}}_f
\end{equation}
where $\phi_s^{\text{SM}} = -(2.1\pm 0.1)^\circ$ \cite{Art15}. Here the process dependence of $\phi_f$ enters via $\Delta \psi_f$, which in the SM is given by doubly Cabibbo-suppressed penguin topologies (see Sec.~\ref{sec:phidandphis}). %, which will be discussed in detail in Sec.~\ref{sec:phidandphis}.
We may now rewrite Eq.~\eqref{eq:asl} as 
\begin{equation}\label{eq:aslin}
	a_{\rm{sl}}^s = \left[ (0.46 \pm 0.04) \times 10^{-2} \right] \times \tan\left( \left\langle \phi_s \right\rangle + \Delta \Psi \right) \ ,
\end{equation}
where we have used the measurements of $\Delta\Gamma_s$ and $\Delta M_s$ in Eq.~\eqref{eq:GammaMs}. We emphasize that the numerical suppression in Eq.~\eqref{eq:aslin} already renders the value of $a_{\rm{sl}}^s $ in the range of the current LHCb measurement \cite{LHCb-asl}. 
										
In Eq.~\eqref{eq:aslin}, we used Eq.~\eqref{eq:phisfdef} to define $\phi_s =  \left\langle \phi_s \right\rangle + \Delta \Psi $ to make the link to the exclusive $\phi_s^f$ determinations explicit. Here $\left\langle \phi_s \right\rangle =-(1.5 \pm 1.8)^\circ$ is the average of the different available exclusive $\phi_s^f$ determinations \cite{KKV17}. We emphasize that currently all these determinations are consistent with the SM, which significantly constrains possible new physics effects. To quantify this, we introduced in addition the phase $\Delta \Psi$
\begin{equation}
\Delta \Psi = \rm{arg} \Big[ \sum_f \eta_f w_f e^{i(\phi_s^f - \left\langle \phi_s \right\rangle)}\Big] \ , 
\end{equation}
where $\eta_f$ is the CP eigenvalue of the final state and $w_f$ is a measurable weight function
\begin{equation}
w_f = \Gamma(B_s^0 \to f) \sqrt{ \frac{1- {A}_{\rm{CP}}^{\rm{dir}}(B_s^0\to f)}{1+ {A}_{\rm{CP}}^{\rm{dir}}(B_s^0\to f)}}  \ .
\end{equation}
\begin{figure}[tbp] 
   \centering
  \includegraphics[width=2.6in]{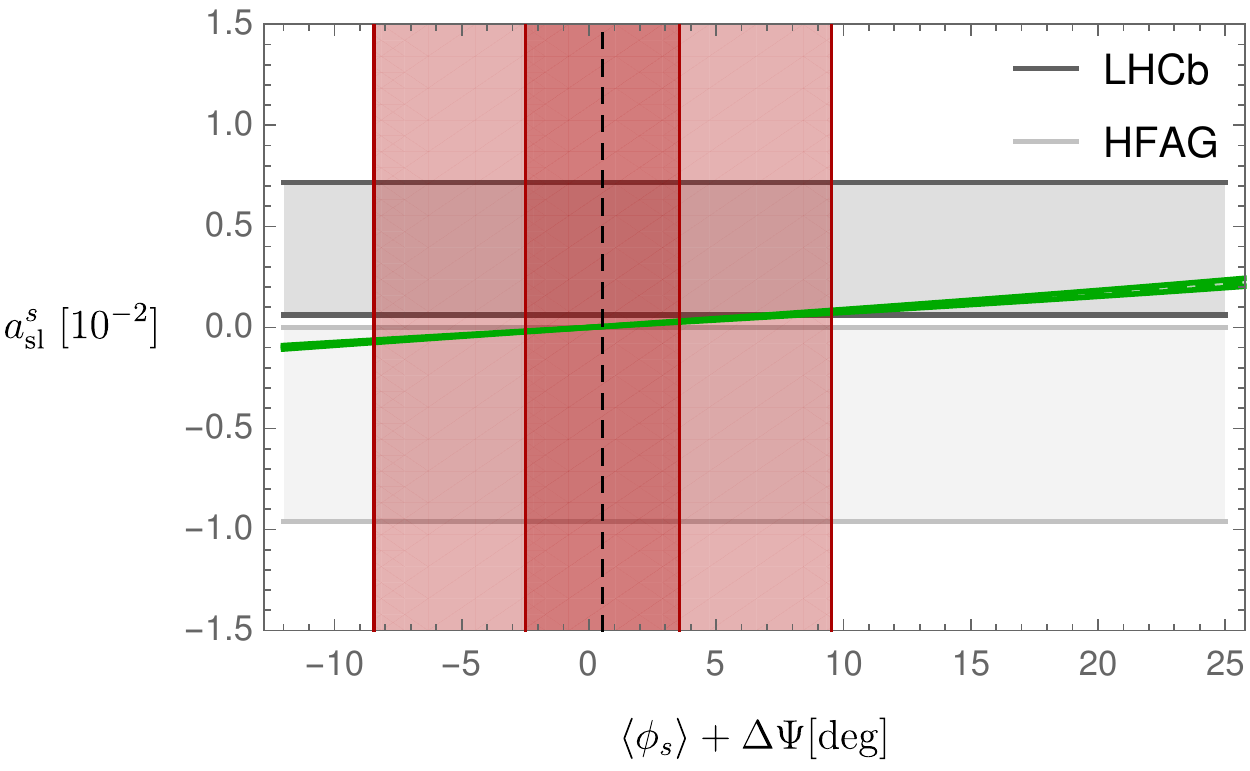} 
  \vspace*{-0.4truecm}
  \caption{Dependence of $a_{\rm sl}^s$ on $\langle\phi_s\rangle+\Delta\Psi$ from (\ref{eq:aslin}). The broad vertical bands correspond to the experimental range in (\ref{asl-num-1}), while the narrow band presents an update scenario. The horizontal bands show the recent LHCb determination \cite{LHCb-asl} and HFAG average \cite{HFAG}, respectively.}\label{fig:asl}
\end{figure}													
For the available exclusive measurements we obtain $\Delta\Psi=(2.1 \pm 9.0)^\circ$, which yields \cite{KKV17}
\begin{equation}\label{asl-num-1}
a_{\rm sl}^s = (0.004\pm 0.075)\times10^{-2}.
\end{equation}
In Fig.~\ref{fig:asl}, we show this exclusive constraint combined with the sensitivity of $a_{\rm sl}^s$ on $\langle\phi_s\rangle+\Delta\Psi$ from (\ref{eq:aslin}) and the experimental measurements of $a_{\rm sl}^s$. We find that our analysis significantly limits the size of $a_{\rm sl}^s$ and thereby also the room for new physics. It will be interesting to confront this picture with more precise measurements, in particular for the $B^0_s\to D_s^-D_s^+$ and $B^0_s\to D_s^{*-}D_s^{*+}$ modes, which dominate the current uncertainty of $\Delta\Psi$. To illustrate the effect of such improved measurements, we have added an experimental benchmark scenario in Fig.~\ref{fig:asl}, in which the uncertainty of $\phi_s^{D_s^-D_s^+}$ is reduced by a factor of three.						
Finally, in \cite{KKV17} a new strategy was suggested that exploits the constraints on $a_{sl}^s$ and which opens a new window to search for CP violation in the charm sector. 
													
\section{CP violation in non-leptonic $B$ decays}\label{sec:nonlep}
In the remaining presentation, we focus on CP violation in non-leptonic $B$ decays. These decays are described by an effective field theory, in which heavy degrees of freedom of the SM and possible new particles have been integrated out. 

The SM low-energy effective Hamiltonian for the $B\to f$ decay is given by \cite{Buc95}
\begin{equation}	
\left\langle f | H_{\rm{eff}}| B \right\rangle= \frac{G_F}{\sqrt{2}} \sum_{j=u,c} { V_{jq}^*V_{jb}} \left(\sum_{i=1, 2} {C_i(\mu)}\; \left\langle f |{ {O}_i^{jq}(\mu)}| B \right\rangle + \sum_{i=3}^{10} {C_i(\mu)}\; \left\langle f |{  {O}_i^q } | B \right\rangle\right)
	\end{equation}			
where $V_{ij}$ are the CKM elements. The short-distance contribution to the decay amplitude is described by the  $C_i(\mu)$ Wilson coefficients, which can be calculated in perturbation theory. On the other hand, long-distance physics is described by the matrix elements of the ${O}_i$ operators. Here, the ${O}_1$ and ${O}_2$ are the current-current operators, ${O}_{3,6}$ are the QCD penguin operators and ${O}_{7,\ldots, 10}$ are the electroweak penguin operators. 
The hadronic matrix elements $ \left\langle f |{  {O}_i^q } | B \right\rangle$ can be described in the framework of QCD Factorization (QCDF) \cite{BBNS, BeNe}, perturbative QCD (PQCD) \cite{Li:1994iu}, Soft Collinear Effective Field Theory (SCET) \cite{Bau01} and applications of QCD sum rules \cite{Kho00}. Recently, also two-loop contributions have been studied in QCDF \cite{Ben09, Bel15}. Despite these efforts power corrections that arise in the QCDF description remain difficult to control and in general the description of non-leptonic decays remains a challenge. Alternatively, flavour symmetries can be used to obtain insights into the strong interaction dynamics and its non-perturbative effects. In the following, we focus on strategies that employ flavour symmetries and QCDF to control $SU(3)$ breaking \cite{RF-rev}.
											
In the SM, CP violation is described by the CKM matrix, which is illustrated by a Unitarity Triangle (UT) with the angles $\alpha$, $\beta$ and $\gamma$. A variety of non-leptonic flavour observables can be used to determine the UT parameters, exploiting both the direct and mixing-induced CP asymmetries. Parametrizing a general non-leptonic $B \to f$ decay in the SM as	
	\begin{equation} A(B \to f )  \equiv A_f= e^{i\varphi_1} |A_1| e^{i \theta_1} + e^{i\varphi_2}|A_2| e^{i\theta_2} \ ,
	\end{equation}
	where $\varphi (\theta)$ is represent a weak (strong) phase and similar for the CP conjugate decay $\bar{A}_f$. Direct CP asymmetry is probed via 						
\begin{equation}
{A}_{\rm{CP}}^{\rm{dir}}(B_q\to f)\equiv\frac{|A(B\to f)|^2- |A(\bar{B}\to {f})|^2}{|A(B\to f)|^2+ |A(\bar{B}\to {f})|^2} = \frac{2 |{A}_1| |{A}_2| \sin(\Delta \theta ) \sin\Delta\varphi}{|{A}_1|^2 + |{A}_2|^2 + 2|{A}_1|| {A}_2| \cos(\Delta\theta) \cos\Delta\varphi} \ ,
\end{equation}									
which is governed through the interference between two different decay amplitudes. Direct CP violation, thus requires both a non-trivial CP-conserving strong phase difference $\Delta \theta$ and a non-trivial CP-violating weak phase difference $\Delta\varphi$. 
For neutral $B^0_q$ decays, the $B_q^0$-$\bar{B}^0_q$ oscillations give rise to a time-dependent decay rate asymmetry:								
\begin{equation}	
A_{\rm{CP}}(t)  \equiv \frac{\Gamma(B_q^0(t) \to f)-\Gamma(\bar{B}_q^0(t) \to \bar{f})}{\Gamma(B_q^0(t) \to f)+\Gamma(\bar{B}_q^0(t) \to \bar{f})} =				\frac{A_{\rm{CP}}^{\rm{dir}}%(B_q \to f)							
\cos(\Delta M_qt)+A_{\rm{CP}}^{\rm{mix}}%(B_q \to f)
\sin(\Delta M_qt)}{\cosh(\Delta \Gamma_q t/2) +A^{\Delta \Gamma_q} \sinh(\Delta \Gamma_q t/2)} \ ,
\end{equation}
where 
 \begin{equation} 
A_{\rm{CP}}^{\rm{dir}}(B_q^0 \to f)
\equiv \frac{1-|\lambda_{f}|^2}{1+|\lambda_{f}|^2} \ ,  \; \;   A^{\Delta\Gamma}(B_q^0 \to f) \equiv \frac{-2 \rm{Re} \lambda_{f}}{1+|\lambda_{f}|^2} \ , \; \; \lambda_f = \frac{q}{p} \frac{\bar{{A}}_f}{{A}_f}
\end{equation}	
Here 
\begin{equation}
A_{\rm{CP}}^{\rm{mix}}%
(B_q^0 \to f)\equiv \frac{-2 \rm{Im} \lambda_{f}}{1+|\lambda_{f}|^2} = \frac{2|\lambda_f|}{1+|\lambda_{f}|^2} \sin\phi_q \ ,
\end{equation}
measures the mixing-induced CP violation. %Exploiting measurements of the CP asymmetries of non-leptonic decays then allows for the extraction of the CP violating phases: the CKM angle $\gamma$ and the mixing-phases $\phi_d$ and $\phi_s$. 
			
\subsection{Determination of $\gamma$ from $B\to D K$ decays}	
\begin{figure}[t]								
\centering
\subfloat[]{\includegraphics[width=0.3\textwidth]{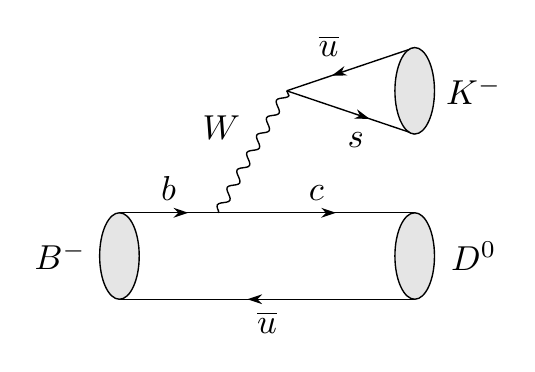}}
\subfloat[]{
\includegraphics[width=0.3\textwidth]{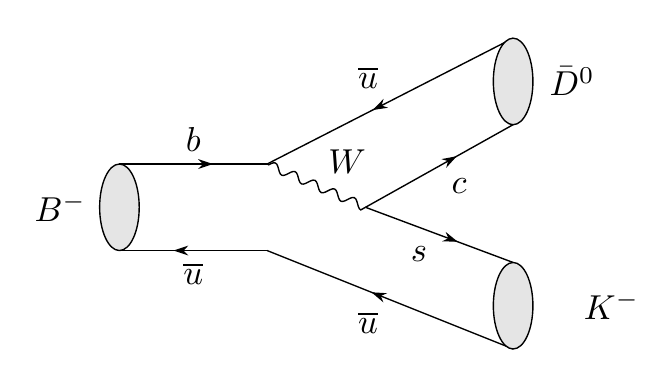}}
\caption{Tree-level contribution to $B^- \to D^0 K^- $ and $B^- \to \bar{D}^0 K^- $ which allow for the determination of the angle $\gamma$. %\propto V_{cb} V_{us(d)}^*$	and  $\propto V_{ub} V_{cs(d)}^*$ 
} 
\label{fig:gammadet}
\end{figure}

The UT angle $\gamma$ is a key input parameter of the CKM matrix and has at the moment, the largest uncertainty. It is given by
\begin{equation}
\gamma = \rm{arg} \left(-\frac{V_{ud} V_{ub}^*}{V_{cd}V_{cb}^*}\right) \ .
\end{equation}
Using $B\to D^{(*)} K^{(*)}$ decays, illustrated in Fig.~\ref{fig:gammadet}, $\gamma$ can be determined in a theoretically clean way. The sensitivity to the angle $\gamma$ comes from the interference between the two different decay topologies with $b\to u \bar{c} s$ and $b\to c \bar{u}s$. For $B^- \to D^0 K^-$ and $B^- \to \bar{D}^0 K^-$, the $D^0$ and $\bar{D}^0$ subsequently decay to the same final state $f$, which gives rise to the interference between the two amplitudes. Several methods to determine $\gamma$ have been proposed \cite{gw, Gro91, ADS, Gir03}. The determination of $\gamma$ is theoretically clean, because these decays are governed by tree-level transitions and in particular no penguin operators contribute. In fact, even electroweak box corrections are tiny \cite{BZ, Bro14}. Due to these favourable features, an experimental precision of $1^\circ$ is expected to be reachable at Belle-II \cite{BelleII} and the LHCb upgrade \cite{LHCb-up}. These exciting prospect make the angle $\gamma$ also an interesting external input parameter in the analyses of non-leptonic decays, as discussed in the following subsections. New physics contributions in the $C_1$ and $C_2$ Wilson coefficients that might influence $\gamma$, have been studied in \cite{BLTXW}. 
											
Finally, also $B_s^0$ decays such as $B_s^0 \to D_s^{\pm} K^{\mp}, ...$ \cite{Ale91, DeB12, RF-BsDsK} provide theoretically clean probes of $\gamma$. Here a time-dependent analysis allows a determination of $\phi_s + \gamma$.  By measuring the CP asymmetries in $B_s^0 \to D_s^\mp K^\pm$, an interesting measurement of $\gamma$ from the $B_s$ system was obtained \cite{LHCb-Bs} 
%\begin{equation}
%\gamma =  (128 ^{+17}_{-22})^\circ \ ,
%\end{equation}	
%which uses the experimental determination of $\phi_s$ from $b\to \bar{c}c$ transitions as an additional input. 
This determination has great potential to be improved in the LHCb upgrade, allowing also to perform a joint analysis to determine $\gamma$ and $\phi_s$ simultaneously  \cite{RF-BsDsK}. Similarly $B_d \to D_s^{\pm} \pi^{\mp}, ...$ decays could be used to probe the combination $\phi_d + \gamma$.

\subsection{The mixing angles $\phi_d$ and $\phi_s$}\label{sec:phidandphis}

As discussed, CP violation from mixing is governed by the mixing phases $\phi_d$ and $\phi_s$. Measurements of these mixing phases play a key role in testing the SM, as new physics might enter. Currently these phases are determined at the few degree and agree with the SM predictions. At Belle-II and the LHC upgrade, determinations of these mixing angles are expected to reach an experimental precision of ${O}(0.5^\circ)$ \cite{BelleII,LHCb-up}, entering a new era of precision physics.  %Therefore, if these phases are affected by new physics, the effect will be small, making it crucial to include higher-order SM corrections. 
In the SM 				
\begin{equation} 
\phi_d^{\rm{SM}}\equiv 2\beta  =2 \rm{arg} \left(-\frac{V_{cd}V_{cb}^*}{V_{td} V_{tb}^*}\right) \ , \;\; 	\phi_s^{\rm{SM}}\equiv 2\beta_s  =2 \rm{arg} \left(-\frac{V_{ts} V_{tb}^*}{V_{cs}V_{cb}^*}\right) \ .
\end{equation}

The golden mode for the determination of $\phi_d$ is  $B_d \to J/\psi K_S$, while for $\phi_s$ the  $B_s \to J/\psi \phi$ decay is most favorable \cite{Fal08}. The CP asymmetries in these decays determine the ``effective'' mixing angle 
\begin{equation}\label{eq:mixingangle}		
\sin\phi_q^{\rm{eff}} = \frac{A_{\rm{CP}}^{\rm{mix}}(B_q^0 \to f)}{\sqrt{1-A_{\rm{CP}}^{\rm{dir}}(B_q^0 \to f)^2}} = \sin\left(\phi_q^{\rm{SM}} + { \Delta \phi_q} + \phi_q^{\rm{NP}}\right) \ ,
\end{equation}							
where the penguin shift { $\Delta \phi_q$} is governed by non-perturbative hadronic parameter rendering it decay mode specific \cite{Fal08}. This term is doubly Cabibbo suppressed, and therefore subleading. Nevertheless, such hadronic effect are mandatory to control in order to differentiate between the SM and NP. This endeavour is complicated by the long distance non-perturbative QCD contributions that enter $\Delta \phi_q$ \cite{Fri15}. Therefore, we focus on a strategy that uses $SU(3)$ symmetry, which provide valuable insights into the hadronic parameters \cite{Fal08, RF99-1, RF99,FFM, Jun12,Ciu05, GL} and in fact enables controlling these penguin effects \cite{DeBrFl}. 
		 
																\subsection{Controlling penguin effects in $B_d^0 \to J/\psi K_S$ and $B^0_s \to J/\psi \phi$}
The penguin shift $\Delta \phi_q$ can be controlled using $U$-spin symmetry of the strong interaction. Assuming only contributions from tree and penguin topologies, we parametrize the golden decay modes $B_d^0 \to J/\psi K_S$ and $B_s^0 \to J/\psi \phi$ as \cite{RF99-1}
\begin{equation}\label{eq:Btofpara}
{A}(B_q^0 \to f) = \left(1 -\frac{\lambda^2}{2}\right) \mathcal{C}' \left[1 + { \epsilon} a'_f e^{i \theta'_f} e^{i\gamma}\right] \ ,
\end{equation}	
where $ \epsilon =\frac{\lambda^2}{1-\lambda^2} \sim 0.05$ and $\lambda = |V_{us}|$ is the CKM element. Here $\mathcal{C}'$ is a CP-conserving hadronic amplitude, $a'$ and the CP-conserving strong phase $\theta'$ parametrize the QCD penguin contributions. Using Eq.~\eqref{eq:Btofpara}, the hadronic phase shift $\Delta\phi_q$ and the CP asymmetries can be expressed in terms of the hadronic penguin parameters. These parameters can then be determined using $U$-spin partner decays in which the penguin effects are not suppressed. For $B_d^0\to J/\psi K_S$ the most prominent candidate is the $B_s^0 \to J/\psi K_S$ decay \cite{RF99}. For the $B_s^0 \to J/\psi \phi$ decay, which decays into two vectors, the transition amplitude and the penguin parameters are polarisation dependent. Suitable penguin control channels are modes with two vector mesons in the final states, most prominently $B_d^0 \to J/\psi \rho^0$, but also $B_s \to J/\psi \bar{K}^{*0}$ \cite{RF99-1, Fal08}. 

In terms of the hadronic parameters, the penguin control decays  are written as
\begin{equation}								
A(B_q \to f) = -\lambda \mathcal{C} \left[1 -  a_f e^{i \theta_f} e^{i\gamma}\right] \ .
\end{equation}	
Here the penguin parameters are not suppressed by the small $\epsilon$. Using now $\gamma$ as external input, the penguin parameters $a$ and $\theta$ can be extracted in a clean way from measurements of the CP asymmetries. In the limit of $SU(3)$ symmetry, $a = a'$ and $\theta = \theta'$. These relations are only affected by non-factorizable $U$-spin breaking corrections \cite{DeBrFl}. 
																\begin{figure}[t]			
\centering
\subfloat[\label{fig:cur}]{\includegraphics[width=0.4\textwidth]{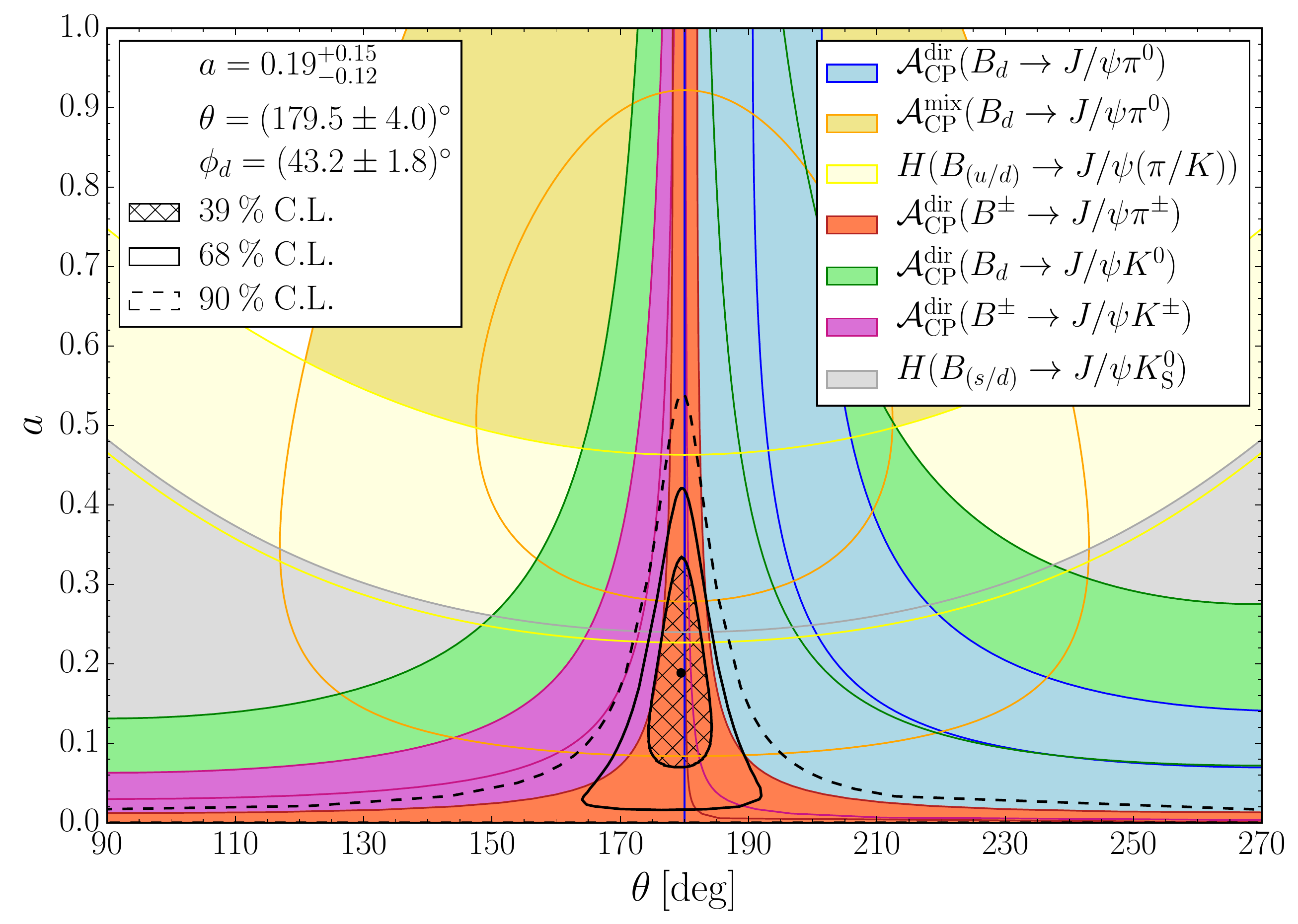}}
\subfloat[\label{fig:benchmark}]{\includegraphics[width=0.43\textwidth]{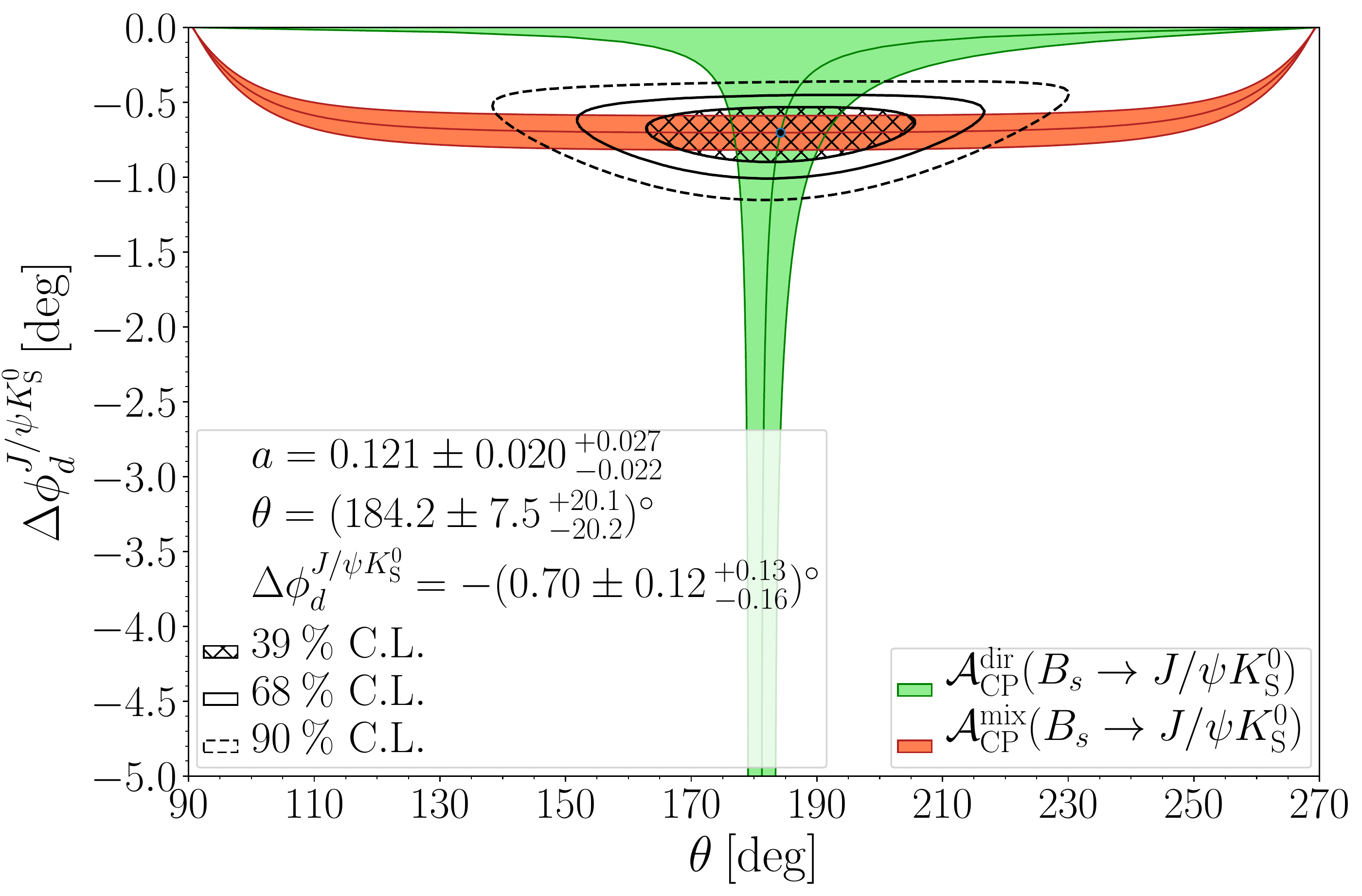}} %was Fig10
\caption{Determination of the penguin parameters $a$ and $\theta$ from current data of $B_q \to J/\psi P$ decays \cite{DeBrFl} and (b) Benchmark scenario illustrating the determination of $\Delta\phi_d$ from the $B_s^0 \to J/\psi K_S$ CP asymmetries (update of \cite{DeBrFl} by K. de Bruyn).} 
\label{fig:phid}
\end{figure}
																Unfortunately, the CP asymmetries of the $B_s^0 \to J/\psi K_S$ have not yet been measured. However, the penguin parameters can be determined when combining all measurements for $B_q\to J/\psi P$ decays as shown in Fig.~\ref{fig:cur}. The extracted penguin parameters result in a penguin shift of (update of \cite{DeBrFl}):
\begin{equation}
\Delta \phi_d^{J/\psi K_S} = (-0.71 ^{+0.56}_{-0.65})^\circ \ ,
 \end{equation}
which relies on some theoretical assumptions. In addition, we show in Fig.~\ref{fig:benchmark} a benchmark scenario for future measurements of the CP asymmetries in $B_s^0 \to J/\psi K_S$ which gives (update of \cite{DeBrFl}):
\begin{equation}
\Delta \phi_d^{J/\psi K_S} = (-0.70 \pm 0.12 (\rm{stat})^{+0.13}_{-0.16} (\text{U-spin}))^\circ \ ,
\end{equation}  						
which is at the same level as the expected experimental precision and shows that the penguin shift is controlled by data.

For the $\phi_s$ determination, the penguin shift can be determined in a similar way using the CP asymmetries in the $B_d^0 \to J/\psi \rho^0$ control channel. In fact, this strategy is already implemented by LHCb. For the different polarizations, $0, \parallel, \perp$, the penguin shifts are found as  \cite{Aaij:2015mea}
\begin{align}
\Delta \phi_{s, 0}^{J/\psi \phi} {}& =0.000^{+0.09}_{-0.011} (\text{stat}) {}^{+0.004}_{-0.009}  (\text{syst}) \text{rad}  \ , \nonumber \\
\Delta \phi_{s, \parallel}^{J/\psi \phi} {}& =0.003^{+0.010}_{-0.014} (\text{stat}) \pm 0.008 (\text{syst}) \text{rad}  \ , \nonumber \\
\Delta \phi_{s, \perp}^{J/\psi \phi} {}& = 0.003^{+0.010}_{-0.014} (\text{stat}) \pm 0.008 (\text{syst}) \text{rad} \ , \nonumber 
\end{align}
which are actually tiny and well under control. Using the extracted information on the hadronic parameters, this strategy also allows for tests of QCD calculations \cite{DeBrFl}. In addition, we note that $\phi_s$ can also be extracted from $B_s^0 \to D_s^-  D_s^+$ using as a control mode the $B_d^0 \to D^- D^+$ decay \cite{Jung:2014jfa, BDD}.   

\subsection{Extraction of $\phi_s$ from $B_s^0\rightarrow K^-K^+$}\label{sec:BstoKK}
It is interesting to compare the determinations of $\gamma$ and $\phi_s$ from tree decays to those from penguin dominated decays, as this sector is in particular sensitive to new heavy particles that might enter in the loops \cite{BuGi}. An interesting decay for the extraction of $\gamma$ and $\phi_s$ is the QCD penguin dominated $B_s^0\to K^- K^+$ decay \cite{RF-rev}. In this strategy, the CP asymmetries play a key role. Using flavour symmetries, the required hadronic inputs can be related to those of the U-spin partner decay $B_d^0\to \pi^-\pi^+$ decay  \cite{RF-99, RF-07, FK, CFMS}. Using the first measurement of CP violation in $B_s^0 \to K^- K^+$, the LHCb collaboration determined \cite{LHCb-BsKK-CP,LHCb-BsKK-gam} 
	\begin{equation}\label{eq:gammaphisLHCb}
\gamma = (63.5^{+ 7.2}_{-6.7})^\circ \quad\quad\quad \phi_s = -(6.9^{+9.2}_{-8.0})^\circ  \ ,
\end{equation}
which agrees with the determinations from pure tree decays. The large uncertainty might be reduced through future data. However, the theoretical precision of this method is limited by $U$-spin breaking corrections to the penguin topologies, making it challenging to reduce the theoretical uncertainty to below ${O}(0.5^\circ)$  \cite{FJV-B, FJV-S}. Therefore, a new strategy was proposed \cite{FJV-B, FJV-S} in which both $\gamma$ and $\phi_d$ are employed as input parameters such that the $SU(3)$-breaking effects can be probed. 

Using the CP asymmetries  $B_s^0 \to K^- K^+$ and  $B_d^0 \to \pi^- \pi^+$, the hadronic parameter $\Delta\phi_s$ in Eq.~\eqref{eq:mixingangle} for the $B_s^0\to K^- K^+$ decay which we define as $\Delta\phi_{KK}$ is determined. Combined with the determination of $\phi_s^{\rm eff}$ from the $B_s^0\to K^- K^+$ CP asymmetries via Eq.~\ref{eq:mixingangle} this then allows for the extraction of $\phi_s$. 
In the determination of $\Delta\phi_{KK}$ both the factorizable and non-factorizable $U$-spin corrections are taken into account. The first are probed using the semileptonic $B_s^0\to K^- \ell^+ \nu_\ell$ and $B_d^0\to \pi^- \ell^+ \nu_\ell$ decays as new ingredients via the ratios
\begin{equation}
R_\pi \equiv \frac{\Gamma(B_d^0 \rightarrow \pi^- \pi^+)}{\left|d\Gamma(B_d^0\rightarrow \pi^- \ell^+ \nu_\ell)/dq^2\right|_{q^2 = m_\pi^2}}		\ , \;\;R_K \equiv \frac{\Gamma(B_s^0 \rightarrow K^- K^+)}{\left|d\Gamma(B_s^0\rightarrow K^- \ell^+ \nu_\ell)/dq^2\right|_{q^2 = m_K^2}}		\ .
\end{equation}
In addition, non-factorizable $U$-spin effects are probed by \cite{FJV-B, FJV-S}
\begin{equation}
			\xi_{\rm{NF}}^a \equiv \left|\frac{a_{\rm{NF}}}{a_{\rm{NF}}^\prime}\right| = \left|\frac{a_{\rm{NF}}^T}{a_{\rm{NF}}^{T \prime}}\right|\left|\frac{1+r_P}{1+r_P^\prime}\right|\left|\frac{1+x}{1+x^\prime}\right| \ ,
\end{equation}
where the primes indicate a $b\to s$ transition such that $\xi_{\rm{NF}}^1$ in the SU(3) limit. Also,  
\begin{equation}
r_P \equiv \frac{P^{(ut)}}{T} \ , \;\;  x \equiv \frac{E+PA^{(ut)}}{T+P^{(ut)}} \ ,
\end{equation}%\sim {O}(\lambda) 
are ratios of tree (T), penguin (P), exchange (E) and penguin-annihilation (PA) topologies. 

The parameter $\xi_{\rm{NF}}^a$ has a theoretically favourable and robust structure in terms of $U$-spin-breaking parameters and allows the use of data to quantify $U$-spin breaking corrections. The tree-level contributions  $a_{\rm{NF}}^T$ were calculated in QCDF \cite{Ben09}. Allowing for $U$-spin breaking of $20 \% $ gives a correction of ${O}(1\%)$. 	The hadronic parameters $r_P$ and $x$ can be determined from data and are ${O}(0.2)$ \cite{FJV-B}. Allowing for $U$-spin breaking of $20 \% $ gives a correction of ${O}(4\%)$ to the ratio $\xi_{\rm{NF}}^a$. The effects of $U$-spin breaking can be probed directly from data. Specifically, future measurements of CP asymmetries of the pure P decays $B_d^0\to K^0\bar{K}^0, B_s^0\to K^0 \bar{K}^0$, would allow for a determination of both $r_P$ and $r_P'$, while the pure E and PA decays  $B_d^0\rightarrow K^+K^-, B_s^0\rightarrow \pi^+\pi^-$ would probe $x$ and $x'$. Finally, combining these corrections leads to $\xi_{\rm{NF}}^a \sim {O}(5 \%)$ \cite{FJV-B, FJV-S}. 

\begin{figure}[t]
\centering
\includegraphics[width = 0.45\linewidth]{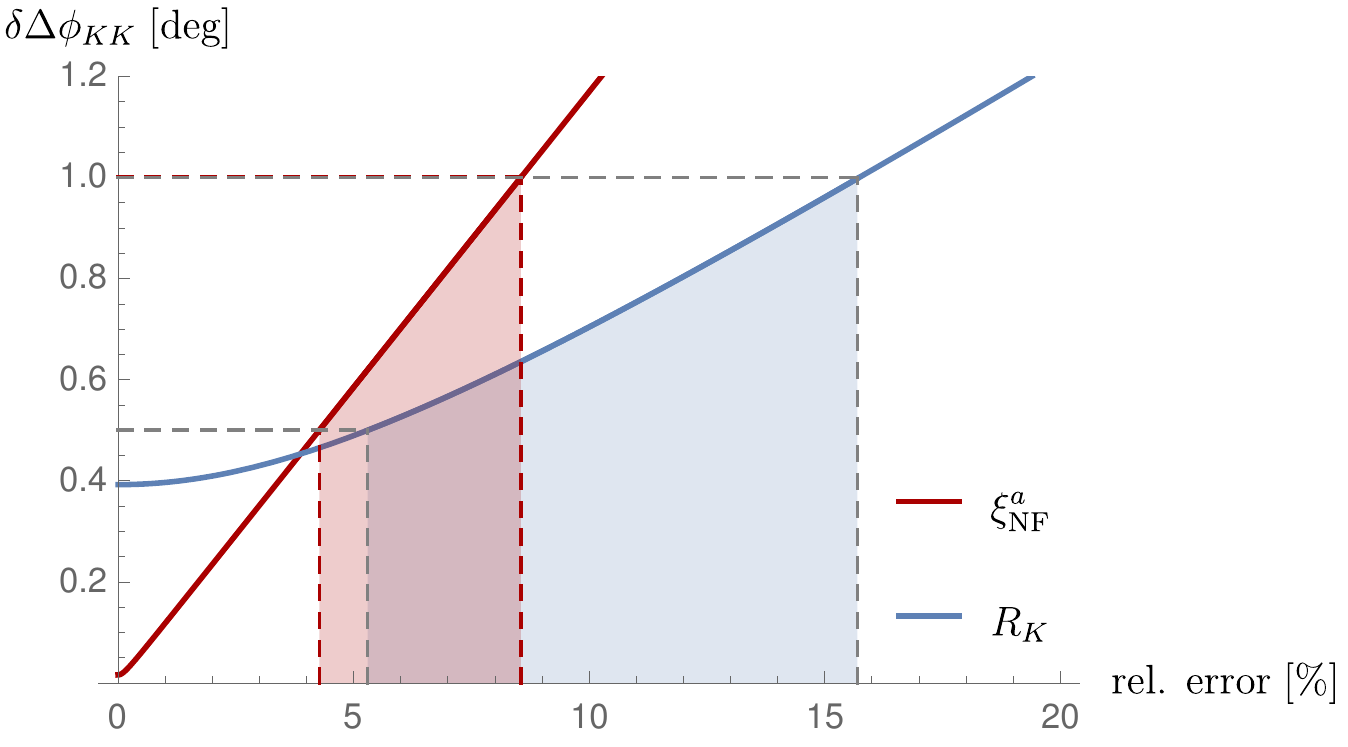}
\caption{Illustration of the error on $\Delta \phi_{KK}$ for the relative error on both the theoretical determination of $\xi^a_{\text{NF}}$ and the experimental semileptonic ratio $R_K$  \cite{FJV-B, FJV-S}. }\label{fig:Rk}
\end{figure}

Figure~\ref{fig:Rk} illustrates the uncertainty on $\Delta\phi_{KK}$ from both the semileptonic ratio $R_K$ and the ratio $\xi_{\rm NF}^a$. Matching the expected experimental precision of $0.5^\circ$ in the upgrade era, would require both a $5\%$ precision on differential rate of $B_s^0\to K^- \ell^+ \nu_\ell$ and a $5\%$ precision on the $SU(3)$-breaking corrections probed by $\xi_{\rm{NF}}^a$. Unfortunately, measurements of the differential decay rate of $B_s^0\rightarrow K^-\ell^+\nu_\ell$ are not available. They are strongly encouraged in order to apply the new strategy. Comparing $\phi_s$ obtained from the penguin-dominated $B_s^0 \to K^- K^+$ decays to the SM predictions and determinations from tree decays as discussed in Sec.~\ref{sec:phidandphis} might reveal new sources of CP violation. 

\section{Search for new physics in $B\to \pi K$}
The $B\to \pi K$ decays are particularly interesting to test the SM, especially to probe possible new physics effects in the electroweak (EW) penguin sector. These decays have been in the spotlight for decades as previous data showed puzzling patterns (see e.g. \cite{Bur04, Bur04a, FRS, Fle08, BDLRR}). 

The $B\to \pi K$ modes are QCD penguin dominated, as the tree amplitudes are $V_{ub}$ suppressed. The EW penguins are colour-suppressed for the $B_d^0 \to \pi^- K^+$ and $B^+\to \pi^+ K^0$ decays, while for $B_d^0 \to \pi^0 K_S$ and $B^+\to \pi^0 K^+$ they are colour-allowed and contribute at the same level as the tree topology \cite{RF-96}. 

The EW penguins are parametrized by	
\begin{equation}		
q e^{i\phi} e^{i\omega} \equiv - \left(\frac{\hat P_{EW}' + \hat P_{EW}'^{\rm C}}{\hat{T}' +\hat{C}'} \right)	 \ ,
\end{equation}
where $\phi(\omega)$ is a CP-violating (conserving) phase. The phase $\omega$ is small and vanishes in the $SU(3)$ limit \cite{NR}. Here $P'_{EW} (\hat{T}')$ and $P_{EW}'^{\rm C} (\hat{C}')$ denote the colour-allowed and colour-suppressed EW penguin (tree) amplitudes. In the SM, these EW penguin parameters can be calculated using $SU(3)$ flavour symmetry to the hadronic matrix elements \cite{NR, BF-98, RF-95}, yielding \cite{Fle18, Fle18a} 
\begin{equation}
q_{\rm{SM}}= (0.68\pm 0.05)R_q\ , \quad \phi_{\rm{SM}}=0^\circ \ ,
\end{equation}
where $R_q$ may differ from 1 through $SU(3)$ breaking corrections. Therefore, a non-zero phase $\phi$ would be a ``smoking-gun'' signal for new CP violating physics. Recently, we pointed out that the current $B_d^0 \to \pi^0 K_S$ CP asymmetries are in tension with the SM \cite{Fle18, Fle18a}. %In order to clarify this situation and to reveal the dynamics underlying the EW penguin contributions, we developed a new strategy to determine the EW parameters $q$ and $\phi$. 

Starting by parametrizing the $B\to \pi K$ decays in terms of the hadronic parameters using the isospin symmetry \cite{Bur04}, we observe the following sum rule \cite{gro, GR}:
\begin{align}\label{eq:SR}
\Delta_{\rm SR} &= 	-   {A}^{\rm{dir}}_{\rm{CP}}(B^+\to \pi^0K^+) \frac{2 {\rm {Br}}(B^+\to\pi^0 K^+)}{\rm{Br}(B_d^0\to \pi^- K^+)} \frac{\tau_{B^0}}{\tau_{B^+}} -  {A}^{\rm{dir}}_{\rm{CP}}(B_d^0\to \pi^0 K_{\rm S}) \frac{2  \rm{Br}(B_d^0 \to \pi^0 K^0)}{\rm{Br}(B_d^0 \to \pi^- K^+)} \nonumber \\
& + {A}^{\rm{dir}}_{\rm{CP}}(B_d^0 
\to \pi^- K^+)  +   {A}^{\rm{dir}}_{\rm{CP}}(B^+ \to \pi^+ K^0) \frac{{\rm {Br}}(B^+ \to \pi^+ K^0)}{\rm{Br}(B_d^0 \to \pi^- K^+)} \frac{\tau_{B^0}}{\tau_{B^+}} = 0 +{O}(r_{(\rm{c})}^2) \ ,						
\end{align}
which vanishes in the SM when neglecting hadronic effects of the order of $r_{(\rm{c})}^2\simeq 0.04$ \cite{Fle18, Fle18a}. Using the current experimental data for the $B\to \pi K$ decays \cite{PDG} and 
\begin{equation}\label{eq:Adirexp}
{A}^{\rm{dir}}_{\rm{CP}}(B_d^0 \to \pi^0K^0) = 0.00 \pm 0.13 \ ,
 \end{equation}
we find that this SM null-test is satisfied experimentally, where the dominant uncertainties is given by the experimental measurement in Eq.~\eqref{eq:Adirexp}, which is actually a combination of the BaBar \cite{Aub08} and Belle \cite{Fuj08}  measurements that differ in sign.  Therefore, we may also use the sum rule in Eq.~\eqref{eq:SR}  to predict the direct CP asymmetry in $B_d^0 \to \pi^0K^0$:  %\vspace{0.1cm}
\begin{equation}\label{eq:Adirsumrule}
{A}^{\rm{dir}}_{\rm{CP}}(B_d^0 \to \pi^0K^0) \equiv A_{\rm CP}^{\pi^0 K_S}= -0.14 \pm 0.03 \ .
\end{equation}
\begin{figure}[t]
\centering
\includegraphics[width = 0.35\linewidth]{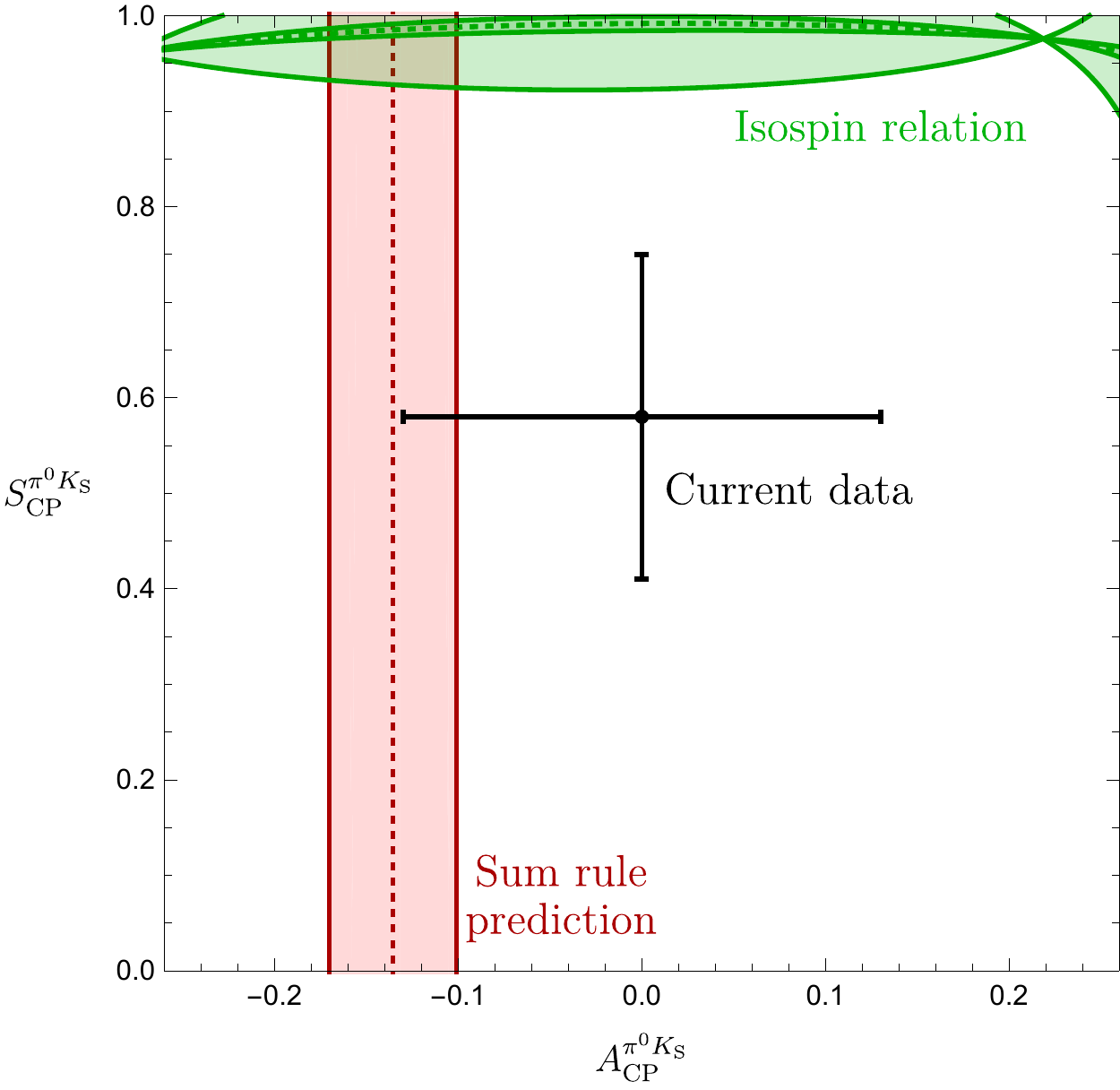}
\caption{Illustration of the $B \to \pi K$ puzzle: the upper green band follows from the isospin analysis while the vertical band shows the sum rule prediction in  Eq.~\eqref{eq:Adirsumrule}. Plot from \cite{Fle18a}.}
\label{fig:puzzle}
\end{figure}
The $B_d^0\to \pi^0 K^0$ decay is of particular interest, because it is the only $B\to\pi K$ mode that exhibits a mixing-induced CP violation. It is given by
\begin{equation}\label{eq:Smix}
		A_{\text{CP}}^{\text{mix}}(B_d^0\to \pi^0 K^0) \equiv S_{\rm CP}^{\pi^0 K_S} = \sin(\phi_d - \phi_{00}) \sqrt{1- \left(A_{\rm CP}^{\pi^0 K_S}\right)^2} \ , 
		 \end{equation}
where $\phi_{00} = \rm{arg}(\bar{A}_{00} A_{00}^*)$ with $A_{00}\equiv A(B_d^0 \to \pi^0 K_S)$ and its CP conjugate decay amplitude $\bar{A}_{00}$ . The $B \to \pi K$ decays obey the isospin amplitude relation 
\cite{NQ, GHLR}
\begin{equation}\label{eq:isospin}
 \sqrt{2} A(B_d^0\to \pi^0 K^0) + 	A(B_d^0\to \pi^- K^+)  = - (\hat{T}' +\hat{C}')\left(e^{i\gamma} - q e^{i\phi}e^{i\omega} \right) = 3A_{3/2} \equiv 3 |A_{3/2}| e^{i \phi_{3/2}}\ ,
\end{equation}
and similar for the CP-conjugate decays. This amplitude relation can be used to determine $\phi_{00}$ via amplitude triangles (see \cite{Fle18a} for a detailed discussion). Using Eq.~\eqref{eq:Smix}, then gives a theoretically clean correlation between the direct and mixing-induced CP asymmetries in $B_d^0\to \pi^0 K_S$  \cite{Fle08, Fle18, Fle18a,GR, groro}. This determination only requires additional information on the normalization of the $I=3/2$ amplitude, which can be obtained using the $SU(3)$ relation \cite{BF-98, GRL}
\begin{equation}\label{eq:tplusc}
|\hat{T}'+\hat{C}'|= { R_{T+C}}\left|\frac{V_{us}}{V_{ud}}\right|\sqrt{2} |A(B^+\to\pi^+ \pi^0)| \ ,
\end{equation}
where $R_{T+C} = 1.2\pm 0.2$ and the uncertainty accounts for non-factorizable $SU(3)$ breaking \cite{BBNS, BeNe}. The obtained correlation for current data is shown in Fig.~\ref{fig:puzzle}, where we also show the sum rule prediction in Eq.~\eqref{eq:SR}. Comparing with the experimental measurements shows a tension that illustrates the puzzling situation in the $B\to \pi K$ system. Moreover, implementing a new constraint that limits the possible triangle amplitude configurations and with the help of the $B\to\pi\pi$ data, this discrepancy can be made even more pronounced \cite{Fle18, Fle18a}. This intriguing situation might indicate NP in the EW penguin sector. 
\begin{figure}[t]								
\centering				
\subfloat[]{\includegraphics[width=0.3\textwidth]{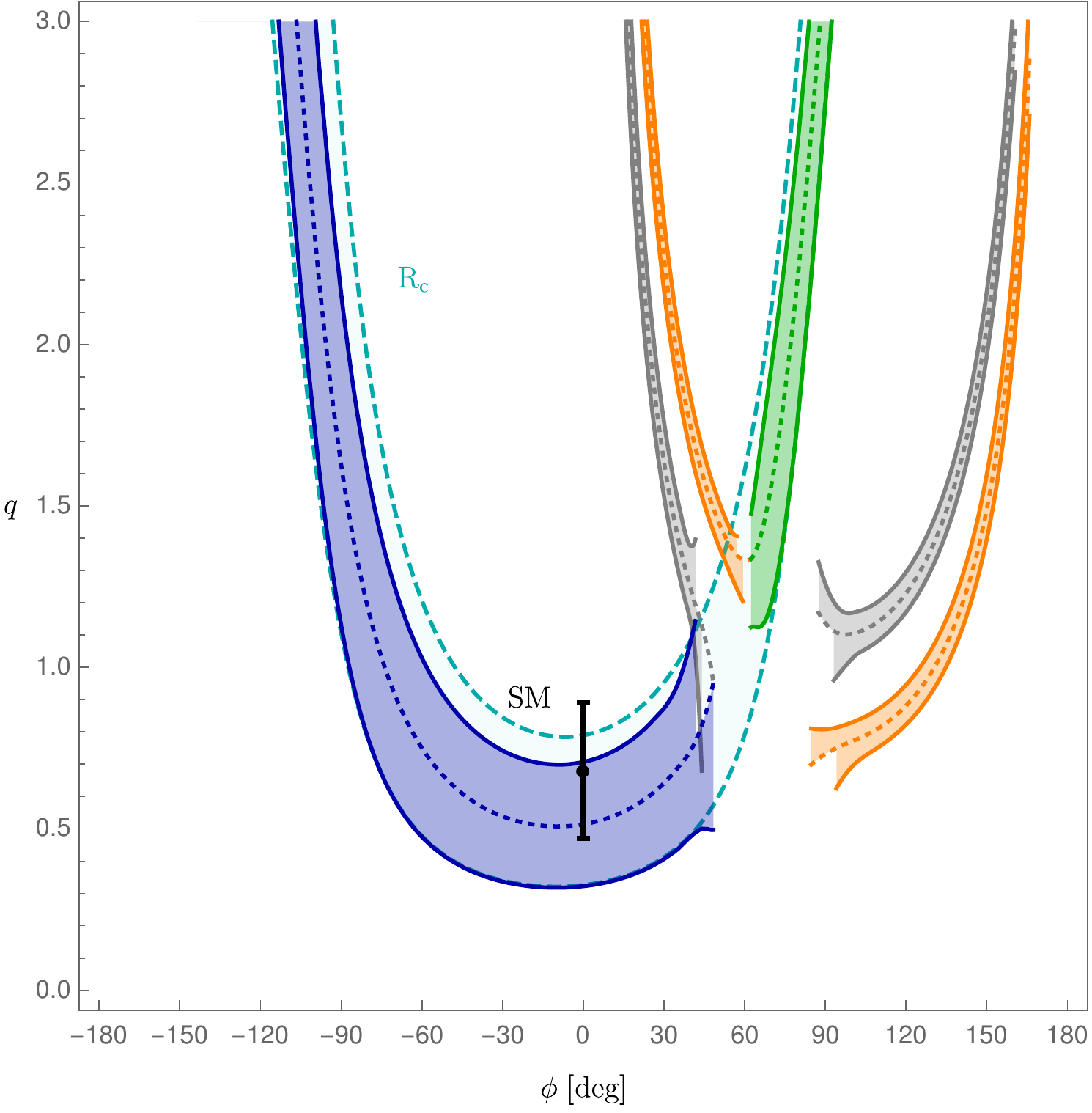}}
\subfloat[]{											\includegraphics[width=0.3\textwidth]{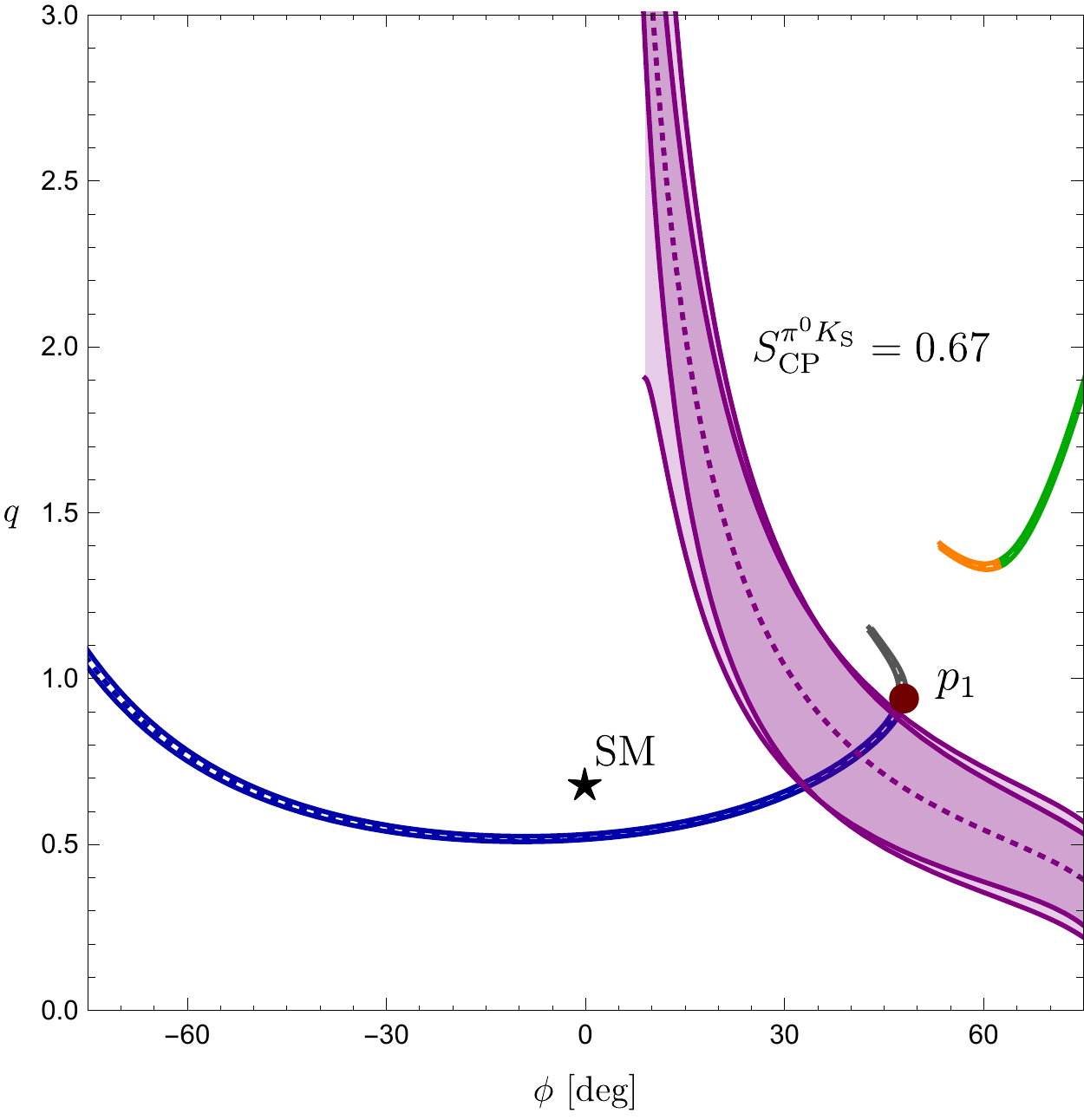}}
\caption{(a) Contours in the $\phi$--$q$ plane for the charged $B\to \pi K $ decays. (b) Future benchmark scenario considering also the mixing-induced CP asymmetry in the $B_d^0 \to \pi^0 K_S$ decays  \cite{Fle18a}. } 
\label{fig:qphiplots}										\end{figure}	
To further study this sector and to reveal the underlying dynamics, it is interesting to determine the electroweak penguin parameters $q$ and $\phi$. A new way to extract them is offered by the analogue of the isospin amplitude relation in Eq.~\eqref{eq:isospin} for the charged $B\to \pi K$ decays, where the experimental data is currently much more precise. The amplitude triangles can be fixed using the $B\to \pi\pi$ data. Using then the measurements of the direct CP asymmetry and the braching ratio of the $B^+\to \pi^+ K^0$ and $B^+\to \pi^0 K^+$ decays, the amplitude triangles allow the determination of the difference $\Delta\phi_{3/2} = \phi_{3/2} - \bar{\phi}_{3/2}$. Finally, with Eq.~\eqref{eq:tplusc} as an input, we can determined $q$ as a function of $\Delta\phi_{3/2}$ and $\phi$ \cite{Fle18a}. The obtained contours in the $\phi-q$ plane are given in Fig.~\ref{fig:qphiplots}. Here the different branches arise because the amplitude triangles obtained through Eq.~\eqref{eq:isospin} have a four-fold ambiguity. We note that there is still a lot of room for NP. In the future, the EW parameters $q$ and $\phi$ can actually be determined using in addition the mixing-induced CP asymmetry of $B_d^0 \to \pi^0 K_S$. The corresponding constraint in the $\phi-q$ plane is obtained by writing $S_{\rm CP}^{\pi^0 K_S}$ in terms of the hadronic parameters, which can be determined from the $B\to \pi\pi$ decays including $SU(3)$ breaking corrections (see e.g. \cite{Aaij:2018tfw} for the most recent LHCb measurements). Effects of color-suppressed EW penguin are included and controlled through experimental data ~\cite{Fle18, Fle18a}. To illustrate this, we consider a future measurement of $S_{\rm CP}^{\pi^0 K_S}=0.67$ in Fig.~\ref{fig:qphiplots}. Here we also show the triangle constraints using a benchmark scenario for the theoretical uncertainties. Figure~\ref{fig:qphiplots} shows that this new strategy can determine the EW penguin parameters in a theoretically clean way. This offers exciting prospects for Belle-II  and the LHCb upgrade.

\section{CP violation in multi-body decays}
Finally, we briefly discuss hadronic multi-body decays, which actually constitute a large part of the branching fraction for non-leptonic $B$ decays. These modes have non-trivial kinematics and, therefore, contain much more information on strong phases than two-body decays. Interesting patterns of CP violation were found in the experimental data, especially for $B\to \pi\pi\pi$ decays \cite{LHCb-three}. The theoretical description of these decays is challenging and they have been studied in a large variety of approaches (for some recent studies see e.g. \cite{Dur15, Ded10, Nog15,Bha15, Bed17} and \cite{Vir16} for a recent review).   

\begin{figure}[t]												\centering														\includegraphics[width=0.6\linewidth]{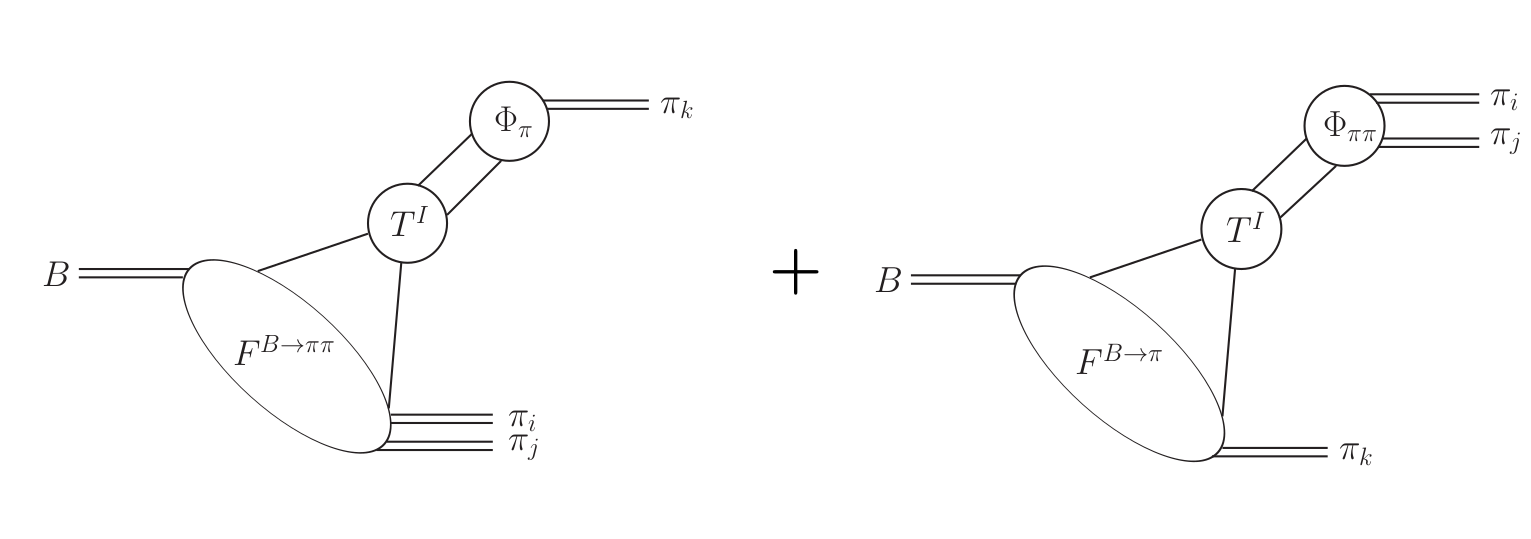}
\caption{Factorization formula at the edge of the Dalitz plot  \cite{Kra15}.}\label{fig:QCDF}									\end{figure}

Recently, a first attempt was made to study non-leptonic charmless three-body $B$ decays in QCDF \cite{Kra15, KKV17three}. Following \cite{Kra15}, the Dalitz plot distribution can be split into three different regions, where different factorization descriptions apply. In the central region, where all the invariant masses are roughly the same, the amplitude is expected to factorize completely in analogy to two-body decays. However, the $b$-quark is too light for this complete factorization to occur. Therefore, the complete Dalitz plot consists only of edges (i.e. regions where two of the decaying particles move collinearly). Here all the resonances are located and the decays resemble quasi-two body decays. In this region, full factorization breaks down and new non-perturbative quantities need to be introduced, as illustrated in Fig.~\ref{fig:QCDF}.
This factorization theorem is given by \cite{Kra15}
\begin{equation}
		\left\langle \pi^+\pi^+\pi^-|{O}_i| B\right\rangle =T^I_i \otimes F^{B\to \pi^+} \otimes \Phi_{\pi^+\pi^-}													+ T^{II}_i \otimes F^{B\to \pi^+\pi^-} \otimes \Phi_{\pi^+} \ ,
		\end{equation}
where $T_i$ are perturbatively calculable hard scattering kernels, $\phi_\pi$ and $F^{B\to \pi}$ are the pion light-cone distribution amplitude (LCDA) and the $B\to\pi$ form factor which are known from two-body decays. The new elements are the $B\to \pi\pi$ form factor $F^{B\to \pi\pi}$ \cite{Fal13, Boe17, FvDV18} and the $2\pi$-LCDA, which introduce non-perturbative strong phases. The $B\to \pi\pi$ vector form factor was studied using light-cone sum rules \cite{Che17, Che17a}, while the normalization of the $2\pi$-LCDA can be obtained from $e^+e^-$ data. Using this information, a first study of the leading-order contributions to CP violation in $B\to \pi\pi\pi$ using QCDF was performed \cite{KKV17three}. However, the lack of knowledge of the scalar $B\to \pi\pi$ form factor and higher-order corrections currently limits this study. These issues require further investigations and many interesting avenues can still be explored, especially considering the amount of data that will become available. 

\section{Summary}
To fully exploit the upcoming high-precision era in $B$ physics requires continued efforts and synergies between theorists and experimentalists. This challenges theorists to find the cleanest strategies in which theoretical uncertainties are well under control and can be further reduced through future experimental data. In this talk, I discussed the clean determination of the semileptonic asymmetry $a_{\text{sl}}^s$, the UT angle $\gamma$ and the mixing phases $\phi_s$ and $\phi_d$ using flavour symmetries. The $SU(3)$-breaking corrections are controlled using QCDF and experimental inputs. This also allows for an extraction of the hadronic parameters, which give insight into long distance physics, thereby paving the road for future theoretical progress. The new strategies offer interesting prospects for Belle-II and the LHCb upgrade, and show that the uncertainties from strong interaction effects can be sufficiently controlled. Finally, these strategies may either once again confirm the SM or establish new physics.

\section*{Acknowledgments}
I would like to thank Robert Fleischer for the pleasant collaboration and useful discussions. I would also like to thank Kristof de Bruyn for providing updated plots and numerics for Sec.~\ref{sec:phidandphis}, and Ruben Jaarsma, Rebecca Klein, Eleftheria Malami, Thomas Mannel and Javier Virto for the fruitful collaboration on the various topics discussed in this talk. This work is supported by the Deutsche Forschungsgemeinschaft (DFG) within research unit FOR 1873 (QFET).

%%%%%%%%%%%%%%%%%%%%%%%%%%%%%%%%%%%%%%%%%%%%%%%%%%%%%%%%%%

\end{document}